\documentclass[conference]{IEEEtran}
\IEEEoverridecommandlockouts
\usepackage{cite}
\usepackage{amsmath,amssymb,amsfonts}
\usepackage{gensymb}
\usepackage{algorithmic}
\usepackage{graphicx}
\usepackage{textcomp}
\usepackage{xcolor}
\usepackage{multirow}
\usepackage{svg}
\usepackage{caption}
\usepackage{subcaption}
\usepackage{float}
\usepackage{url}
\usepackage{tikz}
\usepackage{pgfplots}
\newcommand\copyrighttext{%
  \footnotesize \textcopyright 2025 IEEE. Personal use of this material is permitted. 
  Permission from IEEE must be obtained for all other uses, in any current or future 
  media, including reprinting/republishing this material for advertising or promotional 
  purposes, creating new collective works, for resale or redistribution to servers or 
  lists, or reuse of any copyrighted component of this work in other works.}

\newcommand\copyrightnotice{%
\begin{tikzpicture}[remember picture,overlay]
\node[anchor=south,yshift=10pt] at (current page.south) 
{\fbox{\parbox{\dimexpr\textwidth-\fboxsep-\fboxrule\relax}{\copyrighttext}}};
\end{tikzpicture}%
}

\pgfplotsset{compat=1.17}
\usetikzlibrary{shapes.geometric, arrows}
\def\BibTeX{{\rm B\kern-.05em{\sc i\kern-.025em b}\kern-.08em
    T\kern-.1667em\lower.7ex\hbox{E}\kern-.125emX}}
\begin{document}

\title{SMCNet: Supervised Surface Material Classification Using mmWave Radar IQ Signals and Complex-valued CNNs\\

\thanks{The authors acknowledge the financial support by the Federal Ministry of Education and Research of Germany in the programme of “Souverän. Digital. Vernetzt.”. Joint project 6G-life, project identification number: 16KISK002}
}

\author{\IEEEauthorblockN{Stefan Hägele, Fabian Seguel, Driton Salihu, Adam Misik, and Eckehard Steinbach}
\IEEEauthorblockA{\textit{Technical University of Munich} \\
\textit{School of Computation, Information and Technology}\\
\textit{Chair of Media Technology}\\
\textit{Munich Institute of Robotics and Machine Intelligence}\\
\{stefan.haegele, fabian.seguel, driton.salihu, adam.misik, eckehard.steinbach\}@tum.de}
}

\maketitle
\copyrightnotice

\begin{abstract}
Understanding surface material properties is crucial for enhancing indoor robot perception and indoor digital twinning. However, not all sensor modalities typically employed for this task are capable of reliably capturing detailed surface material characteristics.
By analyzing the reflected RF signal from a mmWave radar sensor, it is possible to extract information about the reflective material and its composition from a certain surface.
We introduce a mmWave MIMO FMCW radar-based surface material classifier SMCNet, employing a complex-valued Convolutional Neural Network (CNN) and complex radar IQ signal input for classifying indoor surface materials.
While current radar-based material estimation approaches rely on a fixed sensing distance and constrained setups, our approach incorporates a setup with multiple sensing distances.
We trained SMCNet using data from three distinct distances and subsequently tested it on these distances, as well as on two more unseen distances.
We reached an overall accuracy of 99.12-99.53\% on our test set.
Notably, range FFT pre-processing improved accuracy on unknown distances from 25.25\% to 58.81\% without re-training.
\end{abstract}

\begin{IEEEkeywords}
mmWave radar, classification, materials, indoor environments, signal processing, machine learning, deep learning, digital twinning
\end{IEEEkeywords}

\section{Introduction}
In recent years, compact radar sensors have seen growing demand due to their widespread availability, small form factor, and robust sensing performance across various applications.
Currently, mmWave radars in frequency range from 30-300 GHz are frequently employed in automotive for target classification \cite{automotive1}, often in conjunction with camera or LiDAR sensors \cite{phillip, mengchen, fusion1, fusion2}.
Indoors, mmWave radars are mainly used for tracking, localization and occupancy detection \cite{tracking, localization, occupation}.
\\
Robot perception and digital twinning of indoor environments can be significantly improved by incorporating material information of walls and object surfaces into a virtual representation of the environment.
To achieve these enhancements, obtaining accurate estimates of the surface materials is essential.
Given different reflection properties of various materials, mmWave radar can be used to estimate surface materials by detecting variations in the reflected RF signal or in single scattering parameters \cite{material1, material2, material3, material4}.
However, this estimate must be robust to environment changes and indoor distortions such as interference, noise, and multi-path reflections.
Deep learning approaches for radar signal processing are capable of extracting important features from perceived scenes, making them robust against the aforementioned distortion effects. 
\\
Hence, we propose a supervised surface material classifier SMCNet for indoor material classification. SMCNet has the capability to process complex numbers and utilizes complex-valued radar signals as input. This is adjusted to the radar IQ signal's complex nature and eliminates the need to transfer the complex radar signal to the real- or image-like domain, potentially losing features and correspondences of the signal.
Image-based approaches relying on radar images are particularly susceptible to this, as they require transforming the radar signal into specific radar maps such as Range-Doppler-, Range-Angle-, or Cross-Range map.
We evaluate the complex radar signal as two different input versions. The first is the analog-to-digital converted (ADC) IQ signal directly provided by the radar, while the second is a range FFT-transformed version of the same signal.
We utilize the complex-valued Convolutional Neural Network (CNN) model introduced in \cite{fabian} to process the complex signal input and evaluate our system's performance with different sensor distances to the measured surface.
This is in contrast to material classification in \cite{material1, material2, material3}, which use fixed setups along with real-valued deep learning to measure surface materials.
Furthermore, complex networks are known to process complex signals in a more effective manner \cite{compNN}.
In the end, we want to provide a more general way to measure and classify indoor surface materials.
Our contributions are as follows:
\begin{itemize}
    \item We collect a radar ADC IQ signal dataset consisting of different indoor surfaces (5 classes, 5 distances) using a $20\times20$ MIMO FMCW mmWave imaging radar.
    \item We design a complex-valued CNN for surface material classification with complex radar signals as input.
    \item We evaluate our model with different pre-processing methods and against other classification models.
\end{itemize}
Section \ref{meth} introduces the data acquisition, its partitioning for training and testing, and the signal pre-processing along with the SMCNet design. After that, the results, evaluation and conclusion of our proposed system are presented in Section \ref{res} and Section \ref{dis}.
\section{Proposed System and Methodology}
\label{meth}
\subsection{Data Collection}
\begin{figure}
  \begin{subfigure}{0.225\textwidth}
    \fbox{\includegraphics[width=\linewidth]{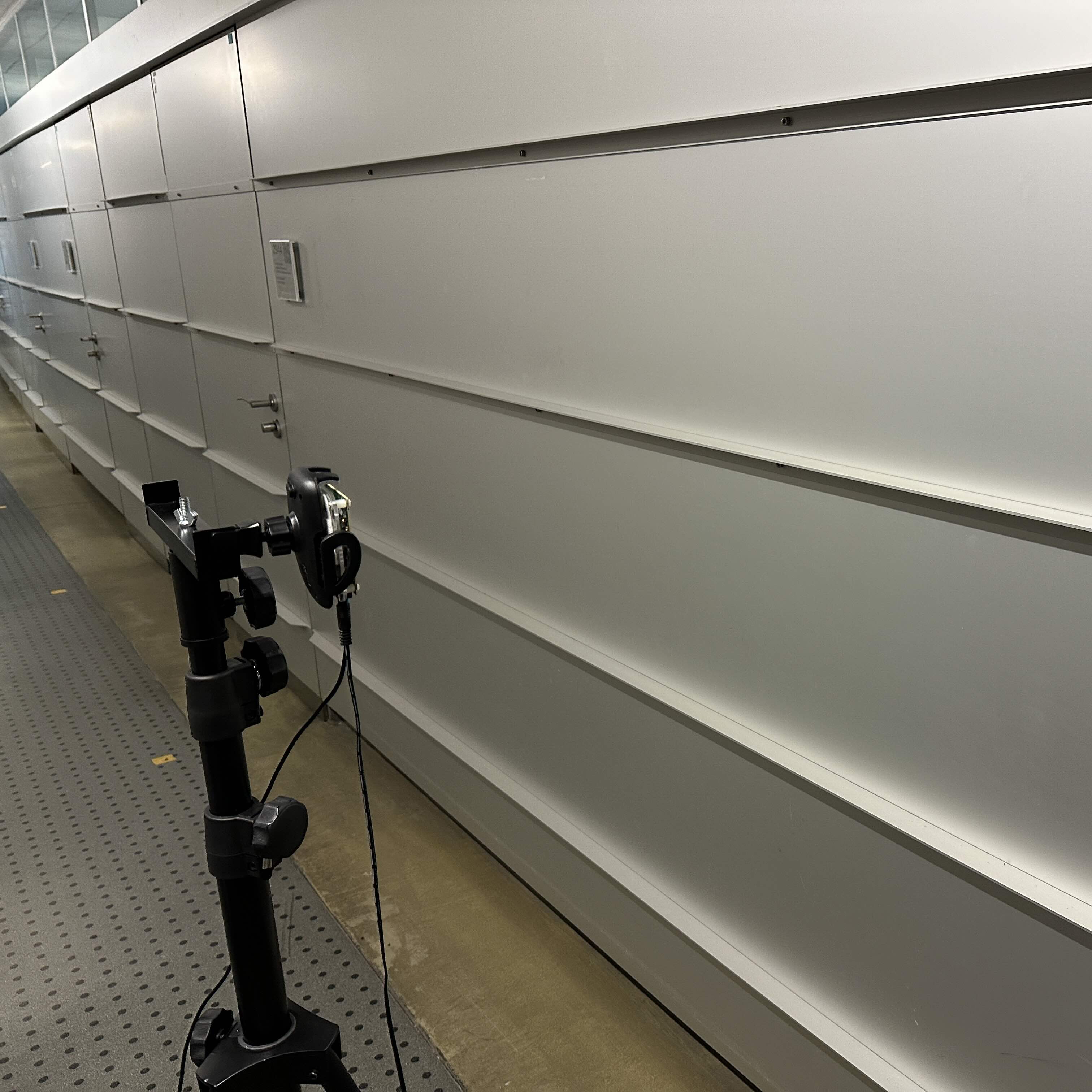}}
  \end{subfigure}
  \hspace{0.001\textwidth}
  \begin{subfigure}{0.225\textwidth}
    \fbox{\includegraphics[width=\linewidth]{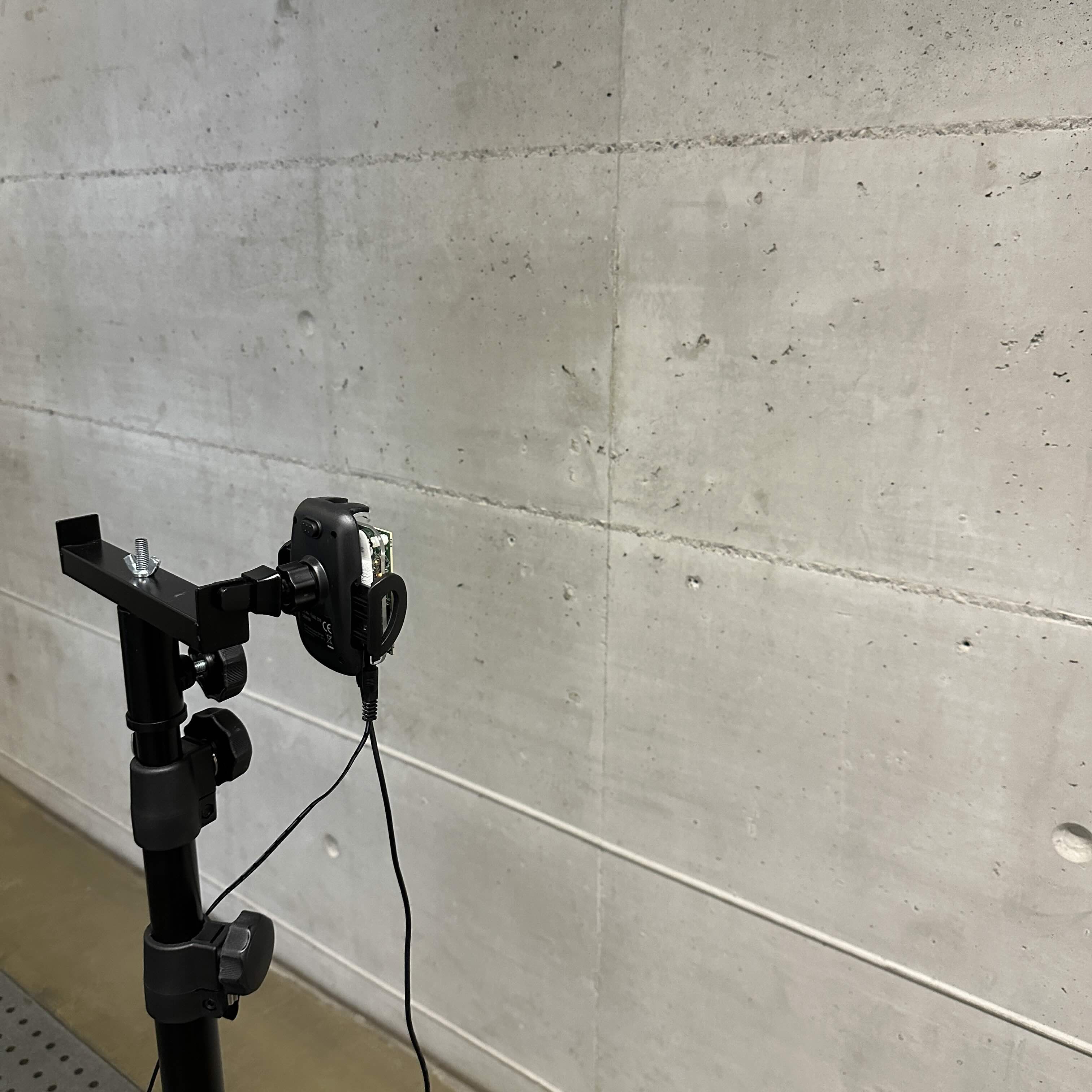}}
  \end{subfigure}
  \caption{Data collection of different surfaces with different sensor distances in an indoor environment.}
  \label{testbed}
\end{figure}
\begin{figure}
    \centering
\begin{tikzpicture}
\pgfmathsetmacro{\cubex}{3}
\pgfmathsetmacro{\cubey}{3}
\pgfmathsetmacro{\cubez}{6}

\definecolor{aliceblue} {rgb}{0.94,
0.97, 1.0}

\draw[black,fill=aliceblue] (0,0,0) -- ++(-\cubex,0,0) -- ++(0,-\cubey,0) --++(\cubex,0,0) -- cycle;
\draw[black,fill=aliceblue] (0,0,0) -- ++(0,0,-\cubez) -- ++(0,-\cubey,0) -- ++(0,0,\cubez) -- cycle;
\draw[black,fill=aliceblue] (0,0,0) -- ++(-\cubex,0,0) -- ++(0,0,-\cubez) -- ++(\cubex,0,0) -- cycle;

\draw [decorate, decoration={brace,amplitude=15pt,raise=1pt}] (-3,-3,0) -- ++(0,\cubey,0) node [midway, anchor= east, xshift=-4mm, outer sep=10pt]{$Rx$-$Tx$ channels};

\draw [decorate, decoration={brace,amplitude=15pt,raise=1pt}] (-3,0,0) -- ++(0,0,-\cubez) node [midway, anchor= south east, outer sep=10pt]{$N$ time samples};

\draw[dashed] (-0.5,0,0) --++ (0,-3,0);
\draw[dashed] (-1,0,0) --++ (0,-3,0);
\draw[dashed] (-1.5,0,0) --++ (0,-3,0);
\draw[dashed] (-2,0,0) --++ (0,-3,0);
\draw[dashed] (-2.5,0,0) --++ (0,-3,0);

\draw[dashed] (0,-0.5,0) --++ (-3,0,0);
\draw[dashed] (0,-1,0) --++ (-3,0,0);
\draw[dashed] (0,-1.5,0) --++ (-3,0,0);
\draw[dashed] (0,-2,0) --++ (-3,0,0);
\draw[dashed] (0,-2.50,0) --++ (-3,0,0);

\draw[dashed] (0,-0.5,0) --++ (0,0,-6);
\draw[dashed] (0,-1,0) --++ (0,0,-6);
\draw[dashed] (0,-1.5,0) --++ (0,0,-6);
\draw[dashed] (0,-2,0) --++ (0,0,-6);
\draw[dashed] (0,-2.50,0) --++ (0,0,-6);

\draw[dashed] (-0.5,0,0) --++ (0,0,-6);
\draw[dashed] (-1,0,0) --++ (0,0,-6);
\draw[dashed] (-1.5,0,0) --++ (0,0,-6);
\draw[dashed] (-2,0,0) --++ (0,0,-6);
\draw[dashed] (-2.5,0,0) --++ (0,0,-6);

\draw[dashed] (0,0,-1) --++ (0,-3,0);
\draw[dashed] (0,0,-2) --++ (0,-3,0);
\draw[dashed] (0,0,-3) --++ (0,-3,0);
\draw[dashed] (0,0,-4) --++ (0,-3,0);
\draw[dashed] (0,0,-5) --++ (0,-3,0);

\draw[dashed] (0,0,-1) --++(-3,0,0);
\draw[dashed] (0,0,-2) --++(-3,0,0);
\draw[dashed] (0,0,-3) --++(-3,0,0);
\draw[dashed] (0,0,-4) --++(-3,0,0);
\draw[dashed] (0,0,-5) --++(-3,0,0);

\end{tikzpicture}
    \caption{Qualitative visualization of a radar data cube.}
    \label{cube}
\end{figure}
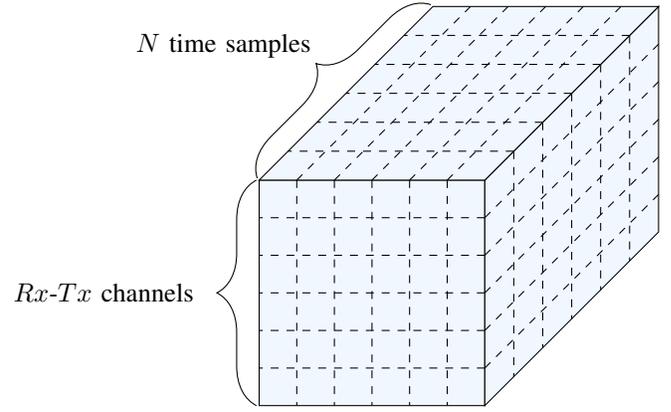

There are only few public datasets available for indoor radar research.
Due to this lack of data, we recorded an indoor material surface dataset consisting of 1360 raw ADC IQ samples structured as radar data cubes for training and testing.
The data was collected in 3 different distances to the measured surface, namely 50 cm, 70 cm, and 100 cm.
This was conducted in 15 different measurement sessions. Despite placing the sensor at the same distance from the surface, small deviations in angular orientation as well as small deviations in distance prevents 1:1 similarity of the measured signal.
The data for testing was also recorded in a separate session from the training data. Hence, multiple measurement sessions ensure higher degree of real-life applicability. 
We collected data for five different surface classes, namely concrete, drywall, glass, metal, wood.
Additionally, 160 out of the 1360 samples were collected at distances of 60 cm and 80 cm for testing only. These samples are used to test  our approach on unknown distances.
The data was collected with the \textit{IMAGEVK-74} 4D imaging radar sensor from \textit{Vayyar} and \textit{minicircuits.com} \cite{vayyar}.
The sensor was configured at a center frequency of $f_{c} =65.5$ GHz and a bandwidth of $B=5$ GHz. The high bandwidth provides a high resolution of the given scene.
A single captured sample (datacube) contains $Rx\times Tx\times N$ complex numbers, while  $Rx=20$, $Tx=20$ is the number of receive- and transmit antennas and $N=100$ is the number of fast time sample points. This creates 400 virtual channels with $N$ entries from every $Rx$-$Tx$ pair which is arranged as a cube. Such cube is depicted in Fig. \ref{cube}.
Each entry represents a complex intermediate frequency (IF) sample point obtained through mixing and down-conversion by the sensor.
One sample cube $\tilde{\boldsymbol{X}}_{j}$ can mathematically be described as:
\begin{equation}
    \tilde{\boldsymbol{X}}_{j} \in \mathbb{C}^{Rx\times Tx\times N}
\end{equation}
This is reshaped to a $(Rx\cdot Tx)\times N$ complex input $\boldsymbol{X}_{j}$ suitable as SMCNet input.
\begin{equation}
    \boldsymbol{X}_{j} \in \mathbb{C}^{(Rx\cdot Tx)\times N}
\end{equation}
For our current application, we use single shot measurements, where no Doppler information (dynamical information) over several pulses is retrieved.

\subsection{Pre-processing}
\begin{figure}
    \centering
    \includegraphics[width=\linewidth]{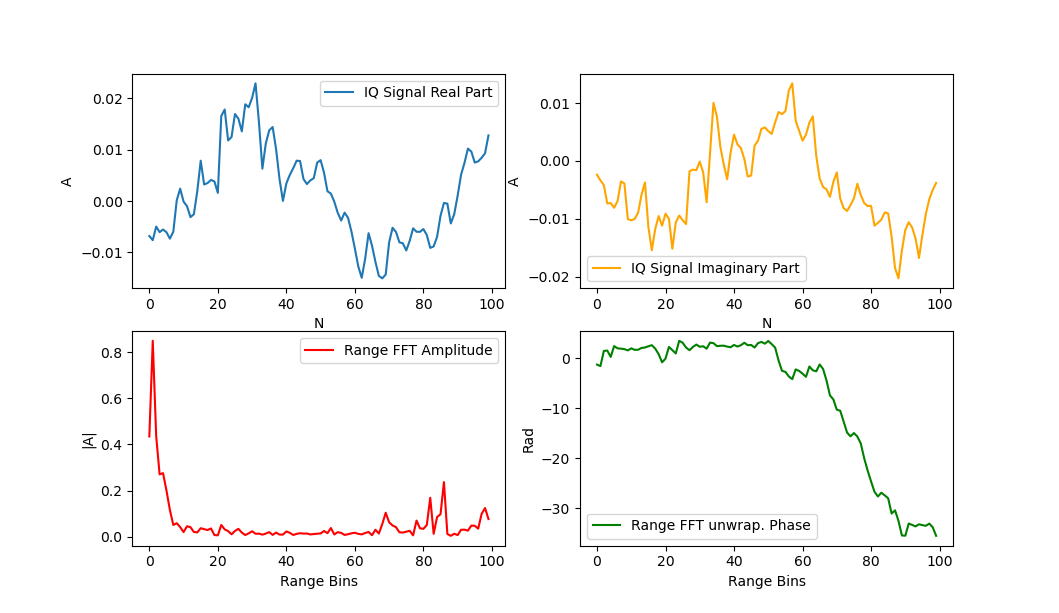}
    \caption{Single channel visualization of an example sample of the collected data (with- and without FFT pre-processing).}
    \label{sample}
\end{figure}

We compare the performance of the proposed method using two different inputs, namely the ADC IQ signal and a range FFT transformed version. Due to the system design of FMCW radars, frequency components contained in the down-converted IQ signal correlate with a certain radar target range and can be obtained by Fourier transform.
Firstly, we use the complex IQ signal directly, without pre-processing.
Secondly, we obtain the range FFT from the complex IQ signal by applying the FFT over each virtual $Rx\text{-}Tx$ channel
\begin{equation}
 X_{i,l}[k] = \sum_{n=0}^{N-1}x_{i,l}[n] e^{-j2\pi \frac{k}{N} n},
\end{equation}
where $i$ is the sample index, $l$ is the channel index within the sample $i$, and $N$ is the number of  points in fast time. This results in range FFT phasor (amplitude and phase) of the transformed IQ signal.
These two different inputs are later fed into our SMCNet.
Fig. \ref{sample} depicts a single channel of one example sample of the input data as IQ signal and as range FFT. Clearly visible is the near field clutter within the range profile on the lower left, followed by the actual reflection pattern of the surface.
\subsection{Neural Network Design}
To process the complex data, we leverage a complex-valued Neural Network architecture. This structure includes complex weights and biases, complex 2D max pooling as well as complex convolutions \cite{fabian}.
\begin{figure}
    \centering
    \includegraphics[width=0.5\textwidth]{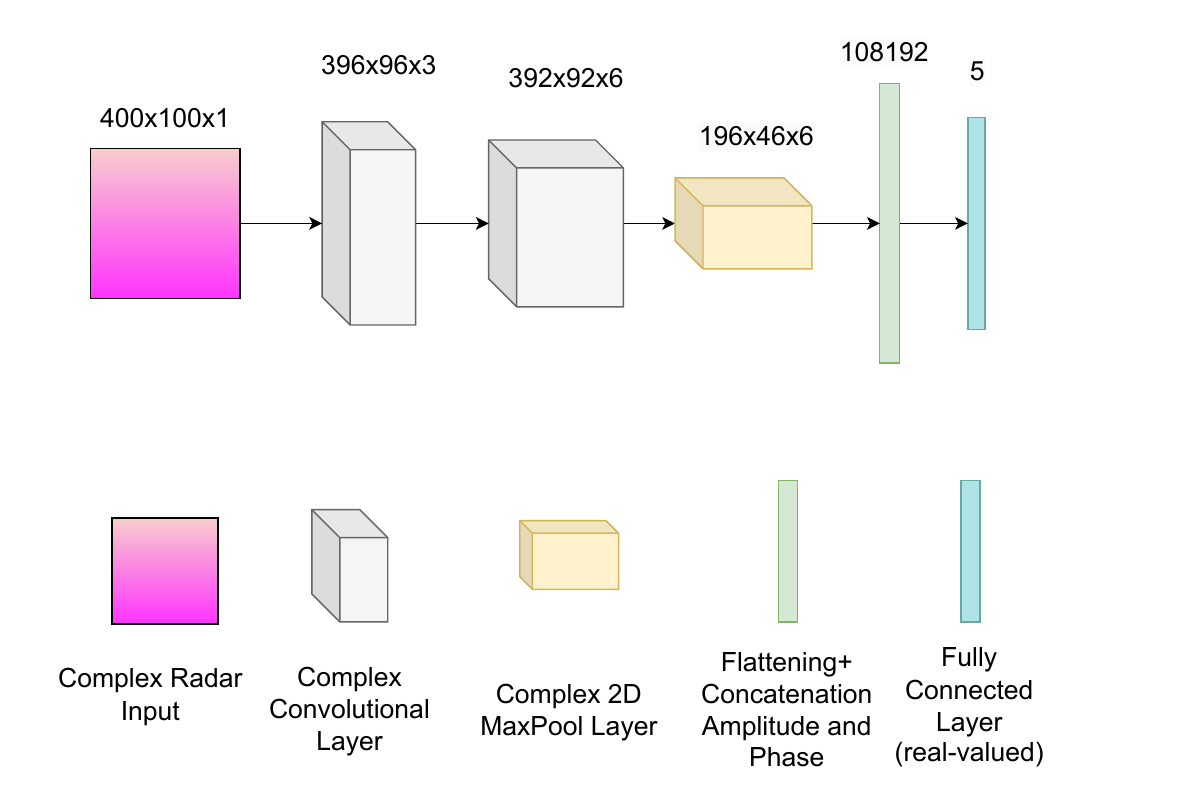}
    \caption{SMCNet structure for processing complex radar signals.}
    \label{network}
\end{figure}
Fig. \ref{network} depicts the structure of the proposed complex network. The network contains two complex convolutional layers with $5\times5$ kernels followed by a single complex max pooling and flattening+concatenation layer.
The concatenation transfers the complex feature vector into a real feature vector by concatenating amplitude and phase.
The real vector is used for the final classification decision of our architecture.
This is needed since the employed cross-entropy loss function $H(\boldsymbol{\hat{y}},\boldsymbol{y})$ in (\ref{crossentropy}) is a real function, rather than a complex one.
\begin{equation}
    H(\boldsymbol{\hat{y}},\boldsymbol{y}) = - \sum_{j=1}^{C} y_{j} \log(\hat{y}_{j})
    \label{crossentropy}
\end{equation}
$C$ is the number of classes, while $\boldsymbol{\hat{y}}$ and $\boldsymbol{y}$ are predicted distribution and ground truth, respectively.
Furthermore, naive complex batch normalization $cBN(.)$ is applied, normalizing real and imaginary part separately along with a batch size of 16.
\begin{equation}
    cBN(z) = BN(Re\{z\}) + j BN(Im\{z\})
\end{equation}
Finally, the training itself is conducted in 10 epochs and with a complex ReLU $cReLU(z)$ activation function \cite{fabian}.
\begin{equation}
    cReLU(z)
    \begin{cases}
    z, &\text{if $Re\{z\}$, $Im\{z\}$} \geq 0\\
    0, &\text{else}
    \end{cases}
\end{equation}
\section{Results and Discussion}
\label{res}
\subsection{SMCNet}
The collected data at 50 cm, 70 cm and 100 cm is uniformly divided into training set and test set. This means that each distance is trained and tested equally. The subsequent testing can be divided into two steps.
The first is conventional testing with unseen data at the three sensor distances 50 cm, 70 cm, and 100 cm.
We compared here the performance of the raw IQ signal and the transformed range FFT.
Fig. \ref{iqtest} displays the confusion matrices of the two approaches on the test set.
\begin{figure}
    \centering
    \scalebox{.475}{\begin{tikzpicture}[scale=1]
    \begin{axis}[
            colormap={bluewhite}{color=(white) rgb255=(255,195,130)},
            xlabel=Prediction,
            xlabel style={yshift=-5pt},
            ylabel=Ground Truth,
            ylabel style={yshift=5pt},
            xticklabels={Drywall, Wood, Concrete, Metal, Glass}, %
            xtick={0,...,4}, %
            xtick style={draw=none},
            yticklabels={Drywall, Wood, Concrete, Metal, Glass}, %
            ytick={0,...,4}, %
            ytick style={draw=none},
            enlargelimits=false,
            xticklabel style={
              rotate=90
            },
            nodes near coords={\pgfmathprintnumber\pgfplotspointmeta},
            nodes near coords style={
                yshift=-7pt
            },
        ]
        \addplot[
            matrix plot,
            mesh/cols=5, %
            point meta=explicit,draw=gray
        ] table [meta=C] {
            x y C
            0 0 100
            1 0 0
            2 0 0
            3 0 0
            4 0 0
            
            0 1 0
            1 1 100
            2 1 0
            3 1 0
            4 1 0
            
            0 2 0
            1 2 0
            2 2 100
            3 2 0
            4 2 0
    
            0 3 1.12
            1 3 0
            2 3 0
            3 3 98.88
            4 3 0
    
            0 4 0
            1 4 1.25
            2 4 0
            3 4 0
            4 4 98.75
            
        }; %
    \end{axis}
\end{tikzpicture}}
    \hfill
    \scalebox{.475}{\begin{tikzpicture}[scale=1]
    \begin{axis}[
            colormap={bluewhite}{color=(white) rgb255=(255,195,130)},
            xlabel=Prediction,
            xlabel style={yshift=-5pt},
            ylabel=Ground Truth,
            ylabel style={yshift=5pt},
            xticklabels={Drywall, Wood, Concrete, Metal, Glass}, %
            xtick={0,...,4}, %
            xtick style={draw=none},
            yticklabels={Drywall, Wood, Concrete, Metal, Glass}, %
            ytick={0,...,4}, %
            ytick style={draw=none},
            enlargelimits=false,
            xticklabel style={
              rotate=90
            },
            nodes near coords={\pgfmathprintnumber\pgfplotspointmeta},
            nodes near coords style={
                yshift=-7pt
            },
        ]
        \addplot[
            matrix plot,
            mesh/cols=5, %
            point meta=explicit,draw=gray
        ] table [meta=C] {
            x y C
            0 0 100
            1 0 0
            2 0 0
            3 0 0
            4 0 0
            
            0 1 0
            1 1 99.79
            2 1 0.21
            3 1 0
            4 1 0
            
            0 2 0
            1 2 4.17
            2 2 95.83
            3 2 0
            4 2 0
    
            0 3 0
            1 3 0
            2 3 0
            3 3 100
            4 3 0
    
            0 4 0
            1 4 0
            2 4 0
            3 4 0
            4 4 100
            
        }; %
    \end{axis}
\end{tikzpicture}}
    \caption{Raw IQ signal (left) and range FFT (right) confusion matrices with 99.53\% and 99.12\% overall accuracy, respectively.}
    \label{iqtest}
\end{figure}
Both achieve high accuracy, namely 99.53\% using the raw IQ signal and 99.12\% using the transformed range FFT. Only few classification mistakes can be spotted. The concrete surface using the range FFT experiences here the highest amount of error, being confused with a wooden surface by a small margin.
In general, however, the network is able to distinguish different surface materials at different sensor distances with high accuracy and does not structurally confuse decreasing signal levels due to increased distance and higher path loss with a different, less reflective surface material.
It is also worth noting that the accuracy of classifying the raw IQ signal seems to be slightly improved compared to the FFT range. Only by a small margin, but one that has been observed through extensive testing. Hence, the features of the raw IQ samples seem to be more similar to each other compared to the features of the transformed samples.
\begin{figure}
    \centering
    \scalebox{.475}{\begin{tikzpicture}[scale=1]
    \begin{axis}[
            colormap={bluewhite}{color=(white) rgb255=(198,233,255)},
            xlabel=Prediction,
            xlabel style={yshift=-5pt},
            ylabel=Ground Truth,
            ylabel style={yshift=5pt},
            xticklabels={Drywall, Wood, Concrete, Metal, Glass}, %
            xtick={0,...,4}, %
            xtick style={draw=none},
            yticklabels={Drywall, Wood, Concrete, Metal, Glass}, %
            ytick={0,...,4}, %
            ytick style={draw=none},
            enlargelimits=false,
            xticklabel style={
              rotate=90
            },
            nodes near coords={\pgfmathprintnumber\pgfplotspointmeta},
            nodes near coords style={
                yshift=-7pt
            },
        ]
        \addplot[
            matrix plot,
            mesh/cols=5, %
            point meta=explicit,draw=gray
        ] table [meta=C] {
            x y C
            0 0 10
            1 0 90
            2 0 0
            3 0 0
            4 0 0
            
            0 1 27.5
            1 1 0
            2 1 72.5
            3 1 80
            4 1 0
            
            0 2 0
            1 2 3.125
            2 2 16.125
            3 2 0
            4 2 80.75
    
            0 3 0
            1 3 0
            2 3 0
            3 3 100
            4 3 0
    
            0 4 19.375
            1 4 13.125
            2 4 60.625
            3 4 6.875
            4 4 0
            
        }; %
    \end{axis}
\end{tikzpicture}}
    \hfill
    \scalebox{.475}{\begin{tikzpicture}[scale=1]
    \begin{axis}[
            colormap={bluewhite}{color=(white) rgb255=(198,233,255)},
            xlabel=Prediction,
            xlabel style={yshift=-5pt},
            ylabel=Ground Truth,
            ylabel style={yshift=5pt},
            xticklabels={Drywall, Wood, Concrete, Metal, Glass}, %
            xtick={0,...,4}, %
            xtick style={draw=none},
            yticklabels={Drywall, Wood, Concrete, Metal, Glass}, %
            ytick={0,...,4}, %
            ytick style={draw=none},
            enlargelimits=false,
            xticklabel style={
              rotate=90
            },
            nodes near coords={\pgfmathprintnumber\pgfplotspointmeta},
            nodes near coords style={
                yshift=-7pt
            },
        ]
        \addplot[
            matrix plot,
            mesh/cols=5, %
            point meta=explicit,draw=gray
        ] table [meta=C] {
            x y C
            0 0 54.69
            1 0 44.06
            2 0 1.25
            3 0 0
            4 0 0
            
            0 1 33.45
            1 1 40.6
            2 1 25.95
            3 1 0
            4 1 0
            
            0 2 0
            1 2 80.94
            2 2 19.06
            3 2 0
            4 2 0
    
            0 3 0
            1 3 0
            2 3 0
            3 3 88.75
            4 3 11.25
    
            0 4 1.25
            1 4 0
            2 4 0
            3 4 7.81
            4 4 90.94
            
        }; %
    \end{axis}
\end{tikzpicture}}
    \caption{Raw IQ signal (left) and range FFT (right) confusion matrices for unknown distances with 25.25\% and 58.82\% overall accuracy, respectively.}
    \label{rptest}
\end{figure}
\\
Considering the second part of testing, Fig. \ref{rptest} shows the resulting confusion matrices when unknown sensor distances (60 cm, 80 cm) from surface to sensor are tested without adaption or re-training of the network. The prediction accuracy using the raw IQ signal drops to 25.25\% and exhibits almost uniform distribution on average, while the transformed range FFT retains 58.81\% of overall accuracy.
This is worth noting, since the classification accuracy using the raw IQ signal slightly outperformed the accuracy retained from the range FFT at known distances before. This cannot be maintained with sufficient accuracy at unknown measurement distances.
While the prediction labels on basis of the raw IQ signal appear to be very random apart from the metal surface, one can observe a narrowing of distribution over all classes when using the range FFT. This effect appears to be the strongest with glass, wood and drywall surfaces while metal slightly drops and concrete stays in a similar range.
Despite having multiple sessions for data collection, this result suggests a worse degree of generalization using the raw IQ signal, as the difference in accuracy deviates to a large extent.
It also suggests that the signal features learnable by our SMCNet are enhanced significantly by applying a range FFT compared to features contained in the raw IQ signal. A range representation of the radar signal can also partly be associated with the penetration depth of the radar signal in terms of the range peak width provided by the range FFT amplitude. This feature appears to be stronger learnable and detectable than any feature of the time-domain IQ signal.
Hence, SMCNet provides larger capability of generalization when using the pre-processed radar input, rather than the IQ signal.

\subsection{Comparison to other Classification Models}

To further evaluate our proposed model, we compare our model and dataset with different ResNet classification models \cite{resnet}, employed in \cite{material3}, as well as a baseline b0 EfficientNet, introduced for image classification \cite{efficientnet}.
A key difference compared to SMCNet is the input data type, as the ResNet and EfficientNet architectures are designed to process image inputs. To fulfill these requirements, we split real and imaginary part of the complex-valued signal and generate a 3-channel pseudo RGB image with a real part channel, an imaginary part channel and a zero padded channel to meet the requirements for the correct input shape.
It is evident that our model is directly suited for processing complex input, while other classification schemes require adaptations to handle complex signals effectively.
The overall accuracy is shown in Table \ref{comp}. $d_0$ refers here to known distances while $d_1$ represents the unknown, untrained distances.
\begin{table}[]
\centering
\caption{Classification accuracy for known distances ($d_0$) and unknown distances ($d_1$) and different inputs in \%.}
\begin{tabular}{ |c|c|c|c|c| } 
\hline
 Model & IQ $d_0$ & FFT $d_0$  & IQ $d_1$ & FFT $d_1$\\
\hline
SMCNet & \textbf{99.53} & \textbf{99.12} & 25.25 & 58.82  \\ 
ResNet-18 & 85.17 & 72.50 & \textbf{39.69} & 56.86\\ 
ResNet-34 & 73.19 & 76.67 & 36.34 & 57.40\\ 
ResNet-50 & 76.53 & 81.04 & 39.02 & 50.09\\ 
EfficientNet b0 & 52.39 & 94.83 & 39.68 & \textbf{64.68}\\ 
\hline
\end{tabular}
\label{comp}
\end{table}
Our proposed method provides a higher overall classification accuracy when tested on original distances. All models struggle with overall generalization on unknown distances, while the EfficientNet provides a small improvement compared to our SMCNet. However, this improvement is not significant enough to make it practical.
Overall, the range FFT enhances feature strength for all architectures.
Despite being excellent classification architectures for images, these models struggle with complex radar signals as input and appear not as capable to sufficiently extract the important features compared to our model.
A generalization on distance to a large extent is not provided with all approaches, which is something to be researched further.
Another advantage of our proposed SMCNet is its model size. The model is significantly smaller than the image classification models, making it easier to embed in hand-held devices or user equipment.
This can be especially beneficial for indoor scanning using mmWave radar.
The number of model parameters is depicted in Table \ref{params}.
\begin{table}[]
\centering
\caption{Number of parameters for  different classification models.}
\begin{tabular}{ |c|c| } 
\hline
 Model & \# of parameters \\
\hline
SMCNet & \textbf{278859} \\ 
ResNet-18 & 11689512 \\ 
ResNet-34 & 21797672 \\ 
ResNet-50 & 25557032 \\
EfficientNet b0 & 5288548 \\
\hline
\end{tabular}
\label{params}
\end{table}
Furthermore, the higher accuracy demonstrates that our proposed method is better at distinguishing deviations in signal strength caused by distance from deviations caused by different material compounds.

\section{Conclusion and Further Work}
\label{dis}
Our proposed material surface classifier demonstrates how a complex-valued CNN can effectively extract material-specific features from complex radar signals, even when measured at different distances. This also proves that the reflected power level is not the only factor in the decision-making process, as the incoming power changes significantly with varying distances.
We further conclude, complex-valued CNNs can easier generalize on radar range FFT signals, rather than raw ADC IQ signals. This effect is expected to be even stronger when tasks with more complexity are investigated. 
We also see a significant improvement in accuracy compared to current architectures used for material classification.
However, the accuracy achieved for the unknown distances is still not within the practical range for realistic application. The goal of our future work is to generalize the approach to achieve a higher distance-independent accuracy.

\end{document}